\documentclass[twocolumn,prd,amssymb,showpacs,10pt]{revtex4}

\usepackage{graphicx}

\usepackage{bm}

\newcommand{\gsim}{\mbox{\raisebox{-1.ex}{$\stackrel
      {\textstyle>}{\textstyle\sim}$}}}
\newcommand{\lsim}{\mbox{\raisebox{-1.ex}{$\stackrel
      {\textstyle<}{\textstyle \sim}$}}}
\newcommand{\MP}{M_{\rm P}}
\newcommand{\beq}{\begin{equation}}
\newcommand{\beqn}{\begin{eqnarray}}
\newcommand{\eeq}{\end{equation}}
\newcommand{\eeqn}{\end{eqnarray}}

\begin{document}

\title{Stretching the Inflaton Potential with Kinetic Energy}
\author{James E. Lidsey}
\affiliation{Astronomy Unit, School of Mathematical Sciences, 
Queen Mary, University of London, Mile End Road,
London, E1 4NS, UK}


\begin{abstract}
Inflation near a maximum of the potential is studied when non-local 
derivative operators are included in the inflaton Lagrangian. 
Such terms can impose additional sources of friction on the field. 
For an arbitrary spacetime geometry, these effects can be quantified in terms 
of a local field theory with a potential whose curvature around the turning
point is strongly suppressed. This implies that a prolonged phase 
of slow-roll inflation can be achieved with potentials that are otherwise 
too steep to drive quasi-exponential expansion. We illustrate this 
mechanism within the context of p-adic string theory. 

\end{abstract}

\vskip 1pc 
\maketitle

Modern cosmological data indicates that the universe is 
spatially flat with a density perturbation spectrum that is 
almost scale-invariant, Gaussian and adiabatic  
on superhorizon scales \cite{WMAP3}. Slow-roll inflation driven by the  
potential energy $W(\varphi )$ of a self-interacting scalar field 
$\varphi$ provides the simplest explanation for these observations
\cite{simplest}. 
Successful inflation requires the potential to respect the flatness conditions 
$\epsilon \ll 1$ and  $|\eta | \ll 1$, where $\epsilon \equiv  
\frac{1}{2} \MP^2(W'/W)^2$, $\eta \equiv \MP^2W''/W$, a prime 
denotes $d/d \varphi$ and $\MP \equiv (8\pi G)^{-1/2} = 2.4\times 10^{18}\, 
{\rm GeV}$. The inflaton then satisfies 
the equation of motion $3H\dot{\varphi} \simeq -W'$ and its 
energy density drives a phase of 
quasi-exponential expansion $\rho \simeq W \simeq 3\MP^2H^2$. 
Furthermore, quantum fluctuations in the 
inflaton generate a Gaussian curvature perturbation with an 
amplitude and spectral index  
\begin{equation}
\label{perturbs}
{\cal{P}}_{{\cal{R}}} = \frac{1}{24\pi^2\MP^4} \frac{W}{\epsilon}, 
\qquad 1-n_s = 6\epsilon -2\eta  .
\end{equation}
Tensor (gravitational waves) fluctuations are also generated during 
inflation with an amplitude $r=16\epsilon$ 
relative to the density perturbations. 

CMB data favours models with $\epsilon \sim 0$ and 
$\eta <0$ \cite{WMAP3}. These conditions arise naturally 
when the inflaton rolls away from a maximum in its potential. 
Models of this type are referred to generically as {\em hilltop scenarios} 
\cite{lofti}. If the 
mass of the inflaton is unsuppressed, a good approximation 
to {\em any} potential in this regime is given by   
\begin{equation}
\label{quadpot}
W (\varphi ) =W_0- \frac{1}{2} \omega^2\varphi^2  ,
\end{equation}
where the constants $\{ W_0 , \omega \}$ specify the energy scale and 
mass of the inflaton, respectively. Although potentials of this form 
arise ubiquitously in particle physics 
models, theory typically predicts that $|\eta | \gsim 1$, 
which makes it difficult to realise slow-roll inflation. 
(See \cite{lythriotto} for a review).  
In any case, the flatness condition $| \eta | \ll 1$ requires  
$\omega^2\MP^2 \ll W_0$, which implies that  
$\varphi_{\rm end}  \sim \sqrt{W_0} /\omega \gg \MP$
at the end of inflation \cite{lindefast}. This immediately raises concerns 
regarding the role and magnitude of non-renormalizable 
contributions to the tree-level potential.

The purpose of the present work is to 
illustrate how these issues can be simultaneously 
resolved by introducing higher-derivative, non-local 
operators into the inflaton Lagrangian. 
In suitable regions of parameter space, we find that such 
terms act as extra sources of friction on the scalar field, 
thereby causing it to slow down sufficiently to drive inflation. More 
specifically, we show that the cosmic 
dynamics induced by the non-local terms is equivalent to that of a 
{\em local} scalar field theory, where the inflaton evolves 
along an effective hilltop potential with $| \eta | \ll 1$.

The motivation for considering higher-order derivatives    
comes directly from string field theory \cite{stringfield}. 
(See \cite{cubicreviews} for reviews). String field theory predicts
that operators of the form $e^{\Box}$ should be included 
at the vertices. These act as kinetic contributions after 
suitable field redefinitions. Exponential operators also arise in the 
$p$-adic string theory \cite{padic}. In this theory, all the amplitudes  
of the lowest state of the string can be determined exactly in 
terms of a non-local effective action. This  
ensures that the theory admits an ultraviolet completion. 

It is well known that ghost degrees of freedom arise 
in any theory which contains a finite number of 
higher-derivative contributions. However, this is not necessarily
the case in theories possessing infinite-order operators.
In particular, exponential factors contain neither poles nor 
zeros at finite momentum and therefore do not introduce ghosts. 
Furthermore, issues related to classical 
(Ostrogradski \cite{ost}) instabilities are not expected to 
be problematic in such theories, since 
the construction of the Ostrogradski Hamiltonian 
requires a highest derivative term in the action \cite{moeller}. 

We consider a general action of the form 
\begin{equation}
\label{action}
S =\int d^4x \sqrt{-g} \left[ \frac{\MP^2}{2} R + 
\gamma^4 \left( \frac{1}{2}
\phi F (\Box ) \phi -V (\phi) \right) \right]   ,
\end{equation}
where $F(\Box )$ is some function of the D'Alembertian operator, 
$\Box \equiv \frac{1}{\sqrt{-g}}
\partial^{\mu} (\sqrt{-g} \partial_{\mu} )$, $\phi$ is a dimensionless 
scalar field and $\gamma$ is a constant with dimensions of mass. 
We will assume implicitly that the functional form of $F (\Box)$
is determined by some (as yet unknown) realistic string theory, 
where the infinite series of higher-dimensional kinetic operators 
has been explicitly calculated. We therefore keep   
$F(\Box)$ unspecified subject only to the weak condition that it is analytic, 
uniformly convergent and admits a Taylor expansion 
$F(\Box) =\sum_{l=0}^{\infty} c_l \Box^l$, 
where all the $c_l$ are real, $\Box^l$ denotes the $\Box$ 
operator applied $l$ times and $\Box^0 \phi \equiv \phi$. 
We will further assume that 
the scalar field is a ${\cal{C}}^{\infty}$ function to ensure continuity.  
These conditions imply that the Taylor series can be integrated term by term 
when it operates on the scalar field. 

Some of the cosmological consequences of specific models of the form 
(\ref{action}) were recently considered in \cite{anupam,khoury,cline,others}.
In what follows, we absorb 
the $l=0$ term in the expansion of $F(\Box)$ 
into the definition of the potential $V(\phi )$. 
The scalar and Einstein field equations are then given by
\begin{equation}
\label{scalareom}
F(\Box ) \phi = \frac{dV}{d\phi}  
\end{equation} 
and $G_{\mu\nu} = \MP^{-2} T_{\mu\nu}$, respectively, where 
\begin{eqnarray}
\label{emtensor}
\frac{1}{\gamma^4}
T_{\mu\nu} = \sum_{l=1}^{\infty} c_l \sum_{j=0}^{l-1} 
\left\{  \left( \partial_{\mu} \Box^j \phi \right)  
\left( \partial_{\nu} \Box^{l-1-j}\phi \right) 
\right.
\nonumber 
\\
\left.
-\frac{1}{2} g_{\mu\nu} \left( \partial_{\alpha}
\Box^j \phi \right) \left( \partial^{\alpha} 
\Box^{l-1-j} \phi \right) \right.
\nonumber
\\
\left. - \frac{1}{2} g_{\mu\nu} \left( \Box^j \phi \right)
\left( \Box^{l-j} \phi \right) 
\right\} 
\nonumber 
\\
+ \frac{1}{2} g_{\mu\nu} \left( \phi \sum_{l=1}^{\infty} 
c_l\Box^l \phi -2V \right)
\end{eqnarray}
defines the energy-momentum tensor of the inflaton 
\cite{tensorref}. 

For a hilltop potential of the form 
\begin{equation}
\label{hilltop}
V(\phi ) =  V_0 - \frac{1}{2} m^2 \phi^2  ,
\end{equation}
where $\{ V_0,m\}$ are dimensionless constants, 
Eq. (\ref{scalareom}) is solved when the inflaton is 
an eigenfunction of the D'Alembertian operator 
\begin{equation}
\label{eigenfunction}
\Box \phi = - \omega^2 \phi ,
\end{equation} 
such that the eigenvalue $\omega^2$ is a solution to the characteristic 
equation 
\begin{equation}
\label{character}
F(- \omega^2  ) = - m^2  .
\end{equation}
Substituting Eq. (\ref{eigenfunction}) 
into the energy-momentum tensor (\ref{emtensor}) then implies that
\begin{equation}
\label{rescaled}
T_{\mu \nu} = \partial_{\mu} \varphi \partial_{\nu}  \varphi
- \frac{1}{2} g_{\mu\nu} \partial_{\alpha} \varphi \partial^{\alpha}
\varphi - \left( \gamma^4 V_0 + \frac{1}{2} \omega^2  \varphi^2 \right)
g_{\mu\nu}  ,
\end{equation}
where $\varphi$ represents a rescaled field 
\begin{equation}
\label{rescaledfield}
\varphi \equiv A \phi ,  \qquad 
A^2= \gamma^4  F^* (- \omega^2)
\end{equation}
and $F^*(-\omega^2) \equiv (dF/d \Box )_{\Box = -\omega^2}$
denotes the derivative of $F(\Box )$ evaluated at $\Box = -\omega^2$.  

Eqs. (\ref{eigenfunction}) and (\ref{rescaled}) represent  
the field equations for a local scalar field 
minimally coupled to Einstein gravity and self-interacting through 
a hilltop potential (\ref{quadpot}), where the mass of the 
field is determined by the eigenvalue $\omega^2$ and $W_0 \equiv \gamma^4 V_0$. 
The action for such a field is 
\begin{equation}
\label{localaction}
S =\int d^4x \sqrt{-g} \left[ \frac{\MP^2}{2} R +
\frac{1}{2} \varphi \Box \varphi -W_0 +\frac{1}{2} \omega^2  \varphi^2 
\right]  .
\end{equation}

Hence, there exists a correspondence between the dynamics of the 
non-local and local theories (\ref{action}) and (\ref{localaction}). 
The origin of this correspondence can be understood 
by inspecting the form of the energy-momentum tensor (\ref{emtensor}). 
In a local field theory (with a general potential) the third and fourth terms 
in Eq. (\ref{emtensor}) cancel. However, in a non-local theory 
with a hilltop potential, the quadratic $\phi$-dependences in the fourth and
fifth terms cancel and only the constant term $W_0$ remains. 
On the other hand, the third term now generates a 
contribution proportional to $\omega^2 \varphi^2$. 
This implies that the non-local operators effectively 
rescale the mass of the field whilst leaving the 
energy scale $W_0$ invariant. Consequently, 
the qualitative shape of the potential remains unaltered, 
although its curvature will be strongly suppressed 
if $\omega^2 \ll m^2 \MP^2$. In this sense, the potential 
is effectively `stretched' by the higher-derivative operators. 

This suggests the possibility of realising slow-roll, hilltop 
inflation in theories where $| \eta | \gsim 1$ in the absence of  
non-local operators. Inflation is still driven by the potential 
energy around the turning point, but the non-local kinetic terms 
source the Hubble parameter via the Friedmann equation. This imposes    
additional friction on the scalar field, causing    
it to evolve away from the turning point 
more slowly than it would do in a conventional scenario. The dynamics 
of the field is such that the energy scale of inflation is maintained at 
$H^2 \simeq W_0 / 3\MP^2$. 
 
This novel dynamical behaviour can be compared to  
the multi-field, assisted inflationary mechanism \cite{assisted}. 
In this scenario, each field continues to roll down its 
respective potential, but feels the enhanced friction arising 
from the other fields that are present. Thus, slow-roll inflation is 
easier to realise. For example, in the case of $M$ uncoupled fields 
of equal mass, the unique late-time attractor of the 
system is the solution where all the fields are equal. 
The dynamics then corresponds to that 
of a single effective field $\phi = \sqrt{M} \phi_i$ 
of the same mass \cite{kanti}. Consequently, inflation 
may arise $(\phi \gsim 1)$ even though each of the individual fields 
is located in a region of its potential that would be too steep to support 
inflation $(\phi_i < 1)$. However, the above scenario 
is fundamentally different to assisted inflation
since it involves only one field and the 
additional friction is provided by kinetic energy rather than the 
potential energy of multiple fields. 

We may illustrate this explicitly by considering the scalar field action 
\begin{equation}
\label{expaction}
S_{\phi} =\gamma^4 \int d^4x \sqrt{-g}
 \left[ -\frac{1}{2} \phi e^{-\alpha \Box}\phi -V_0 
+\frac{1}{2} m^2\phi^2 \right]  ,
\end{equation}
where $\alpha$ is a constant with dimensions 
of $[{\rm mass}]^{-2}$. In the `linear kinetic' regime, 
defined as the region where higher powers of the D'Alembertian operator 
are negligible, the dynamics is equivalent to that of a canonical 
scalar field $\chi \equiv \gamma^2 \sqrt{\alpha} \phi$ with 
a negative mass-squared $m^2_{\rm lin} = (m^2-1)/\alpha$
and `slow-roll' parameter
$\eta_{\rm lin} = -\MP^2 (m^2-1)/(\gamma^4 \alpha V)$.  
However, the higher powers in $\Box$ 
cause the canonical field $\varphi = \gamma^2 \sqrt{\alpha} m 
\phi$ to behave with an effective mass-squared  
\begin{equation}
\label{expmass}
\omega^2 = \frac{1}{\alpha} \ln m^2  ,
\end{equation}
which implies that 
\begin{equation}
\eta = -\frac{\MP^2}{W} \frac{\ln m^2}{\alpha}
= \left( \frac{\ln m^2}{m^2-1} \right)  \eta_{\rm lin} .
\end{equation}
Hence, both the mass-squared of the field and the curvature 
of the potential are suppressed by a factor $(\ln m^2)/m^2$ 
when $m^2 \gg 1$ and slow-roll inflation will be generic in this region of
parameter space even if $|\eta_{\rm lin} | \gsim 1$. 

Furthermore, the field which plays the role of the inflaton, 
$\varphi = \gamma^2 \sqrt{\alpha m^2}\phi = m \chi $, is 
only an effective field in this context. 
The condition that $\varphi_{\rm end} \gg \MP$ 
at the end of inflation can therefore be realised naturally when $m^2 \gg 1$
without requiring $\chi$ to take super-Planckian values. In principle, 
this alleviates problems that may arise regarding the role (or suppression)
of non-renormalizable terms in the tree-level potential for 
$\chi$.

One immediate question that must be addressed is whether inflation of this 
type is able to generate a (nearly) scale-invariant
spectrum of density perturbations. A complete analysis 
of the infinite-order perturbed field equations 
would be required in order to fully resolve this question. Nevertheless, 
in light of the dynamical correspondence that we have established between 
actions (\ref{action}) and (\ref{localaction}), which 
is valid for an arbitrary spacetime metric including
inhomogeneities, it is reasonable to assume as a first 
approximation that the standard, slow-roll results can be applied to the 
minimally coupled field $\varphi$. 

Given this assumption, the conditions near the maximum of the potential, 
$\epsilon \ll | \eta | \ll 1$, imply that the spectral 
index directly constrains the curvature such that 
$1-n_s \simeq 2|\eta|$. More specifically, it follows that 
$\eta \simeq - \MP^2 \omega^2 /W_0$ and 
$\epsilon \simeq \MP^2 \omega^4 \varphi^2/ 2 W_0^2 \simeq 
\frac{1}{2} \eta^2 \varphi^2/\MP^2$, which implies that the energy 
scale and mass of the inflaton are constrained by observations such that 
\begin{eqnarray}
\nonumber 
\frac{W_0}{\MP^4} \simeq \frac{3\pi^2}{2}{\cal{P}}_{{\cal{R}}} r
, \qquad \frac{\omega^2}{\MP^2} \simeq \frac{3\pi^2}{4} 
{\cal{P}}_{{\cal{R}}} r (1-n_s)
\end{eqnarray}
\begin{eqnarray}
\frac{\MP^2 \omega^2}{W_0} \simeq \frac{1-n_s}{2}  .
\end{eqnarray}

On the other hand, the value of the inflaton 
$N$ e-foldings before the end of inflation is given by 
$\varphi_N \simeq \varphi_{\rm end} e^{-| \eta | N}$ and the  
corresponding value of $\epsilon$ is     
\begin{equation}
\label{epsilonN}
\epsilon_N \simeq \frac{(1-n_s)^2}{8} \left( \frac{\varphi_{\rm end}}{\MP}
\right)^2 e^{-(1-n_s)N}  .
\end{equation}
Hence, if inflation ends when  
$\varphi_{\rm end} \sim \sqrt{W_0}/\omega$ 
\cite{lindefast}, the gravitational wave amplitude is maximized for  
$1-n_s \simeq 1/N$. This yields the upper limits 
$r \lsim 1.5 (1-n_s)$,  
$W_0 \lsim 5.6 \times 10^{-8} (1-n_s) \MP^4$ and 
$\omega  \lsim 1.7 \times 10^{-4} (1-n_s) \MP$. 
Current CMB data indicates that ${\cal{P}}_{{\cal{R}}} = 2.5 \times 
10^{-9}$, $n_s = 0.948^{+0.015}_{-0.018}$ and $r< 0.3$ \cite{WMAP3}. 
For the favoured values $N \simeq 60$ and $1-n_s \simeq 0.05$, we find 
that $r \simeq 0.01$, $W_0  \simeq 3.7 \times 10^{-10} \MP^4$  
and $\omega \simeq 3 \times 10^{-6} \MP$. A relatively high gravitational 
wave background is possible in this case 
because $\varphi_{\rm end} \gg \MP$. 

To illustrate the scenario further, we now focus on the theory for the  
$p$-adic string \cite{padic}. In this theory, the 
world-sheet coordinates of the string span the space of $p$-adic numbers.  
The action which yields the amplitudes for the lowest state 
of the string is then given by  
\begin{equation}
\label{p-adicaction}
S=\gamma^4 \int d^4x \sqrt{-g} \left[ -\frac{1}{2} \psi 
e^{-\alpha \Box} \psi + \frac{1}{p+1}
 \psi^{p+1} \right]  ,
\end{equation}
where $p$ is a positive integer, $\alpha \equiv \ln p /(2m_s^2)$, 
$\gamma^4 \equiv (m_s^4/g_s^2)(p^2/(p-1))$, $m_s$ is the string 
mass scale and $g_s$ is the open string coupling. 
The potential of the field is $V=\gamma^4 (\frac{1}{2}\psi^2 
- \frac{1}{p+1} \psi^{p+1} )$. This has a maximum at $\psi =1$, is unbounded 
from below for $|\psi | > 1$,  
and has a zero-energy vacuum at $\psi =0$. In the linear kinetic  
regime, the canonical field is defined by $\chi \equiv 
\gamma^2 \alpha^{1/2} \psi$ 
with an effective mass-squared $m_{\rm lin}^2 = -2m_s^2(p-1) /\ln p$.

When the field is close to the maximum, 
we may define $\psi \equiv 1-\phi$, where $\phi \ll 1$. Expanding action 
(\ref{p-adicaction}) to quadratic order in $\phi$ then implies that 
\begin{equation}
\label{quadaction}
S = \gamma^4 \int d^4x \sqrt{-g} \left[ 
- \frac{1}{2} \phi e^{-\alpha \Box} \phi 
- V_0 + \frac{p}{2} \phi^2 \right]   ,
\end{equation}
where $V_0 \equiv \frac{1}{2}(p-1)/(p+1)$.  
Action (\ref{quadaction}) is of the form given in Eq. (\ref{expaction}) and 
we may deduce immediately that the scalar field equation (\ref{scalareom}) 
is solved by Eqs. (\ref{eigenfunction})-(\ref{character}) 
with $\omega^2 = \ln p /\alpha = 2m_s^2$. Hence, the effective 
mass of the local field $\varphi$ is uniquely determined by the 
string scale and is independent of $p$. 
This rescaling of the mass in terms of the $p$-adic 
string scale was recently identified in Ref. \cite{cline} 
using perturbative techniques in a spatially isotropic cosmology. 
We have established that such a result holds much more generally.  

Our interpretation of $p$-adic  
inflation in terms of a stretched potential should remain valid  
until $\varphi_{\rm end} 
= \gamma^2 \sqrt{\alpha p} \phi_{\rm end} \sim \sqrt{W_0}/\omega$. 
This implies that $\phi_{\rm end} \sim (p\ln p)^{-1/2}$. 
On the other hand, a full numerical calculation has shown 
that inflation is supported by the non-local contributions 
until $\phi \sim p^{-1/2}$ \cite{cline}. In this particular case, 
therefore, our analysis underestimates the total amount of inflation. 
The difference arises because the small-field approximation that leads to 
Eq. (\ref{quadaction}) breaks down before inflation comes to an end. 
Nonetheless, the uncertainty in the value of $\epsilon_N$ 
is approximately $(\ln p)^{-1}$ and this is less than 
an order of magnitude for $p \lsim 10^4$. 

Before concluding, it is interesting to note that non-local 
theories also have implications for inflationary models 
where the inflaton is rolling down a quadratic potential, $V(\chi ) 
= m^2 \chi^2 /2c_1$. A potential of this form represents a
good approximation to more general models 
where the final (observable) stages of inflation occur 
as the field rolls towards a minimum. One problem with these models
is that cosmological scales cross the horizon 
when the canonical field $\chi \simeq \sqrt{4N}\MP \gg \MP$ 
(for $30 \lsim N \lsim 60$) 
and this requires strong suppression of non-renormalizable terms 
in the potential. However, Eq. (\ref{character}) 
admits the solution $F(\omega^2 )=m^2$ for a quadratic potential 
and the effective 
field is therefore given by $\varphi =  (F^* /c_1 )^{1/2} \chi$, where 
$F^* = (dF/d\Box )_{\Box = \omega^2}$. 
It follows that sufficient inflation is achievable 
with $\chi < \MP$ for any non-local contribution satisfying 
$F^* \gsim 4 c_1 N$. For example, this reduces to the constraint 
$m^2 \gsim 4N$ when $F= \exp (\alpha  \Box)$. 

To summarize, we have found that higher-order derivative operators 
in the inflaton Lagrangian can support a prolonged phase of slow-roll 
inflation near a maximum of the potential. 
Such contributions cause the scalar field to 
slow down in such a way that its dynamics is equivalent to that of 
a local field evolving along a qualitatively similar potential. 
The new ingredient is that the curvature of the 
potential away from the turning point is effectively suppressed, 
whereas the height of the maximum 
remains unaltered. Slow-roll inflation can therefore be realised   
with potentials that would ordinarily be too steep to 
support inflation, which is typically the case with potentials arising 
from particle physics. Furthermore, the inflationary   
expansion is driven by an effective scalar  
field which naturally takes super-Planckian values when the 
physical field remains below the Planck scale. 

The mechanism we have outlined can be applied to any model for which 
the mass of the field is not suppressed in the vicinity of the maximum. 
Moreover, it is independent of the spacetime geometry and arises 
for a generic infinite series of higher-order derivatives. 
In particular, no finite truncation of the series need (or should) be made. 
The key requirement is only that Eq. (\ref{character}) 
should admit a solution with sufficiently small eigenvalue $\omega^2$. 
Thus, the framework that we have developed can be applied directly   
to any string (field) theory for which the tower of derivatives  
has been explicitly calculated. 

A number of open issues remain, however. In general, 
one must address the problem of ghosts and related Ostrogradski instabilities
in non-local theories, although it seems that string theory 
is able to avoid these difficulties. A related question concerns 
the lack of a well-defined 
Cauchy problem in a cosmological model involving infinite-order 
time derivatives. These issues were discussed recently in \cite{khoury,cline}. 
Nonetheless, we have established that the general 
field equations do admit an equivalent 
description in terms of a local theory with a well-defined initial value
formulation. It remains to be shown whether such a solution  
represents a late-time attractor of the system. If this turns out 
to be the case, it would provide strong justification for evaluating the 
inflationary perturbation spectrum in terms of a standard, canonical field. 
This would also provide a concrete framework for relating 
the predictions of string field theory directly to cosmological
observations. Finally, our analysis indicates that fine-tuning
issues which arise when developing inflationary models within a 
local field-theoretic framework may be resolved with a 
fully non-local treatment.  

\section*{Acknowledgments}
We thank N. Barnaby, J. Cline and K. Malik for helpful discussions and 
communications.


\begin{thebibliography}{99}

\bibitem{WMAP3}
D. N. Spergel, {\em et al.}, astro-ph/0603449.

\bibitem{simplest}
A. A. Starobinsky, Phys. Lett. B {\bf 91}, 99 (1980); 
A. H. Guth, Phys. Rev. D {\bf 23}, 347 (1981); 
A. Albrecht and P. J. Steinhardt, Phys. Rev. Lett. {\bf 48}, 1220 (1982); 
S. W. Hawking and I. G. Moss, Phys. Lett. B {\bf 110}, 35 (1982); 
A. D. Linde, Phys. Lett. B {\bf 108}, 389 (1982); 
A. D. Linde, Phys. Lett. B {\bf 129}, 177 (1983). 

\bibitem{lofti}
L. Boubekeur and D. H. Lyth, JCAP {\bf 0507}, 010 (2005), hep-ph/0502047.

\bibitem{lythriotto}
D. H. Lyth and A. Riotto, Phys. Rep. {\bf 314}, 1 (1999), 
hep-ph/9807278. 

\bibitem{lindefast}
R. Kallosh, A. Linde, S. Prokushkin, and M. Shmakova, 
Phys. Rev. D {\bf 65}, 105016 (2002), hep-th/0110089.

\bibitem{stringfield}
M. Kaku and K. Kikkawa,
Phys. Rev. D {\bf 10}, 1110 (1974); 
E. Witten, Nucl. Phys. {\bf B268}, 253 (1986). 

\bibitem{cubicreviews}
K. Ohmori, hep-th/0102085; 
I. Ya. Aref'eva, D. M. Belov, A. A. Giryavets, A. S. Koshelev, and 
P. B. Medvedev, hep-th/0111208; 
A. Sen, Int. J. Mod. Phys. {\bf A20}, 5513 (2005), hep-th/0410103.  

\bibitem{padic}
P. G. O. Freund and M. Olson, 
Phys. Lett. {\bf B199}, 186 (1987); 
P. G. O. Freund and E. Witten, Phys. Lett. {\bf B199}, 191 (1987); 
L. Brekke, P. G. O. Freund, M. Olsen, and E. Witten, 
Nucl. Phys. {\bf B302}, 365 (1988). 

\bibitem{ost}
M. Ostrogradski, Mem. Ac. St. Petersbourg VI, {\bf 4}, 385 (1850). 

\bibitem{moeller}
N. Moeller and B. Zwiebach, JHEP {\bf 0210}, 034 (2002), hep-th/0207107. 

\bibitem{anupam}
T. Biswas, A. Mazumdar and W. Siegel, 
JCAP {\bf 0603}, 009 (2006), hep-th/0508194. 

\bibitem{khoury}
J. Khoury, hep-th/0612052.

\bibitem{cline}
N. Barnaby, T. Biswas and J. M. Cline, hep-th/0612230.

\bibitem{others}
I. Ya. Aref'eva and I. V. Volovich, hep-th/0612098; 
A. S. Koshelev, hep-th/0701103; 
I. Ya. Aref'eva, L. V. Joukovskaya and S. Yu. Vernov, hep-th/0701184.

\bibitem{tensorref}
H. Yang, JHEP {\bf 0211}, 007 (2002), hep-th/0209197; 
G. Calcagni, JHEP {\bf 0605}, 012 (2006), hep-th/0512259.

\bibitem{assisted}
A. R.  Liddle, A. Mazumdar and F. E. Schunck, 
Phys. Rev. D {\bf 58}, 061301 (1998), astro-ph/9804177. 

\bibitem{kanti}
P. Kanti and K. A. Olive, 
Phys. Rev. D {\bf 60}, 043502 (1999), hep-ph/9903524. 

\end{thebibliography}
\end{document}